\documentclass[twocolumn,nofootinbib,showpacs,preprintnumbers,superscriptaddress] {revtex4}

\usepackage{amssymb,amsfonts,amsmath,graphicx,float}
\usepackage{color}
\usepackage{enumerate}
\usepackage{braket}
\usepackage{slashed}
\usepackage{physics}
\usepackage{hyperref}

\newcommand{\dis}[1]{\begin{equation}\begin{split}#1\end{split}\end{equation}}
\newcommand{\etal}{et al.\,}

\newcommand{\bfrac}[2]{{\left(\frac{#1}{#2} \right)  }}
\newcommand{\eq}[1]{Eq.~(\ref{#1})}

\newcommand\gev{\,{\rm GeV}}
\newcommand\mev{\,{\rm MeV}}
\newcommand\kev{\,{\rm keV}}

\newcommand\cm{\,{\rm cm}}

\newcommand\kpc{\,{\rm kpc}}

\newcommand\mchi{{m_\chi}}
\newcommand\mzd{{m_{Z_d}}}

\begin{document}

\title{Searching for Boosted Dark Matter mediated by a new Gauge Boson}

\author{Wonsub Cho}
\email{sub526@skku.edu}
 \affiliation{Department of Physics, Sungkyunkwan University,  2066, Seobu-ro, Jangan-gu, Suwon-si, Gyeong Gi-do, 16419 Korea}

\author{Ki-Young Choi}
\email{kiyoungchoi@skku.edu}
 \affiliation{Department of Physics, Sungkyunkwan University,  2066, Seobu-ro, Jangan-gu, Suwon-si, Gyeong Gi-do, 16419 Korea}

\author{Seong Moon Yoo}
\email{castledoor@skku.edu}
 \affiliation{Department of Physics, Sungkyunkwan University,  2066, Seobu-ro, Jangan-gu, Suwon-si, Gyeong Gi-do, 16419 Korea}

\begin{abstract}
We study the  possibility to directly detect the boosted dark matter generated from the scatterings with high energetic cosmic particles such as protons and electrons. As a concrete example, we consider the sub-GeV dark matter mediated by a $U(1)_D$  gauge boson which has a mixing with $U(1)_Y$  gauge boson in the standard model.
The enhanced kinetic energy of the light dark matter  from the collision with the  cosmic rays  can  recoil the target nucleus and electron in the underground direct detection experiments transferring  enough energy to them to be detectable.
We show the impact of boosted dark matter with existing direct detection experiments as well as collider and beam-dump experiments.
\end{abstract}


\preprint{}

\maketitle

\section{Introduction}
The nature of dark matter (DM) is one of the unsolved problems in the astro-particle physics that spans from the small scales of galaxy  to the large scales of the Universe~\cite{DelPopolo:2013qba}.
The underground direct-detection experiment  is one of the ways to search for the non-gravitational nature of DM  and  the sensitivity of the elastic scattering cross section with nucleon now  goes down to $\sigma_{\chi p}\gtrsim 4.1\times 10^{-47}\cm^2$ at $30\gev$ of DM mass~\cite{Aprile:2018dbl}. The constraints on the scattering cross section of DM with electron is $\sigma_{\chi e } \gtrsim 3\times 10^{-38}\cm^2$ at $100\mev$~\cite{Essig:2012yx,Agnese:2018col,Crisler:2018gci}. 

In these studies of the DM direct detection, the DMs are assumed to be non-relativistic with a Mawell-Boltzmann distribution around the Milky Way galaxy with speed around $10^{-3}c$, with the  speed of light $c$. However recently it was noticed that the elastic scattering of  DMs in the Milky Way  with cosmic ray can change the cosmic ray spectra~\cite{Cappiello:2018hsu} and also  boost DM~\cite{Bringmann:2018cvk,Ema:2018bih,Cappiello:2019qsw}. The boosted DM (BDM) can transfer large momentum to the target and make the recoil energy above the detector threshold even with the light DM.
This was used to search for  dark matter in simple models~\cite{Dent:2019krz,Bondarenko:2019vrb,Guo:2020drq,Wang:2019jtk,Su:2020zny}.

In this paper, we apply this novel method to the light DM mediated by a new $U(1)_D$ gauge boson which has a  mixing with $U(1)_Y$ in the Standard Model~\cite{Babu:1997st,Pospelov:2007mp,Langacker:2008yv},
which is one of the simplest extension of the Standard Model (SM). In this model, the mixing connects the visible and hidden sector 
through the mediation of the gauge bosons and opens the portal to  the DM in the hidden sector. Here the DMs can interact with both nuclei and electrons, and  therefore it is necessary to consider both scatterings with nuclei and electrons in the BDM generation as well as in the direct detection.  This gives  different behavior and constraints compared to  the previous analysis assuming a single kind of interaction.
In this study, we give the realization of the up-scattered DM by cosmic rays of a vector-mediation~\cite{Dent:2019krz} and complements the  existing constraints on this model~\cite{Chun:2010ve,Foot:2014uba,Foot:2014osa,Foot:2016wvj,Choi:2017kzp,Dutra:2018gmv,Li:2019drx,Bernal:2019uqr,Emken:2019tni,Ibe:2019gpv}.

In Sec.~\ref{model}, we introduce the model we consider, and in Sec.~\ref{BDM} we summarize the generation of BDM and attenuation. In Sec.~\ref{BDMdetection}, we show the results with constraints from BDM, and conclude in Sec.~\ref{conclusion}.

\medskip

\begin{figure*}[!t]
\begin{center}
\begin{tabular}{cc} 
 \includegraphics[width=0.4\textwidth]{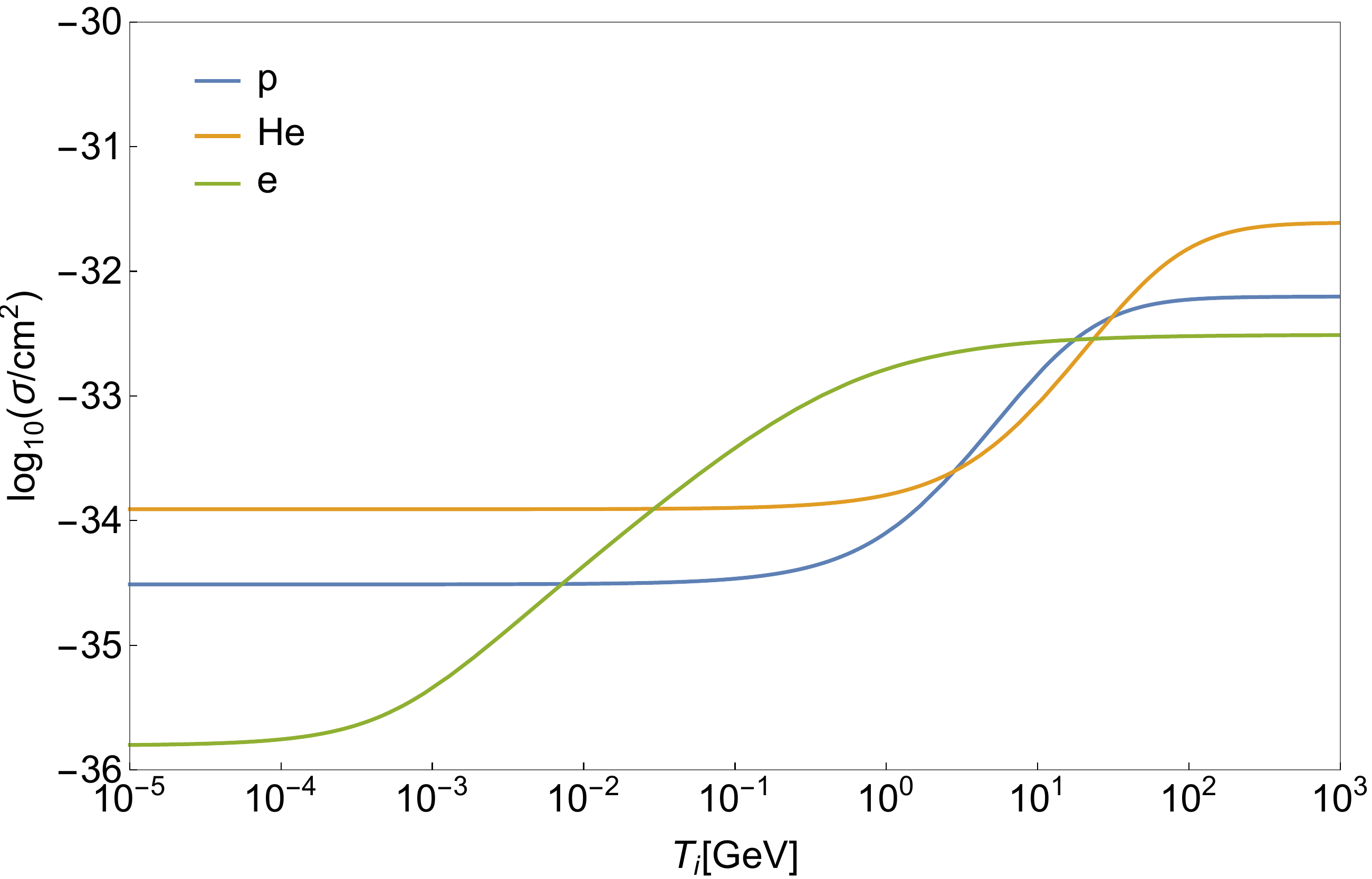}
 &
 \includegraphics[width=0.4\textwidth]{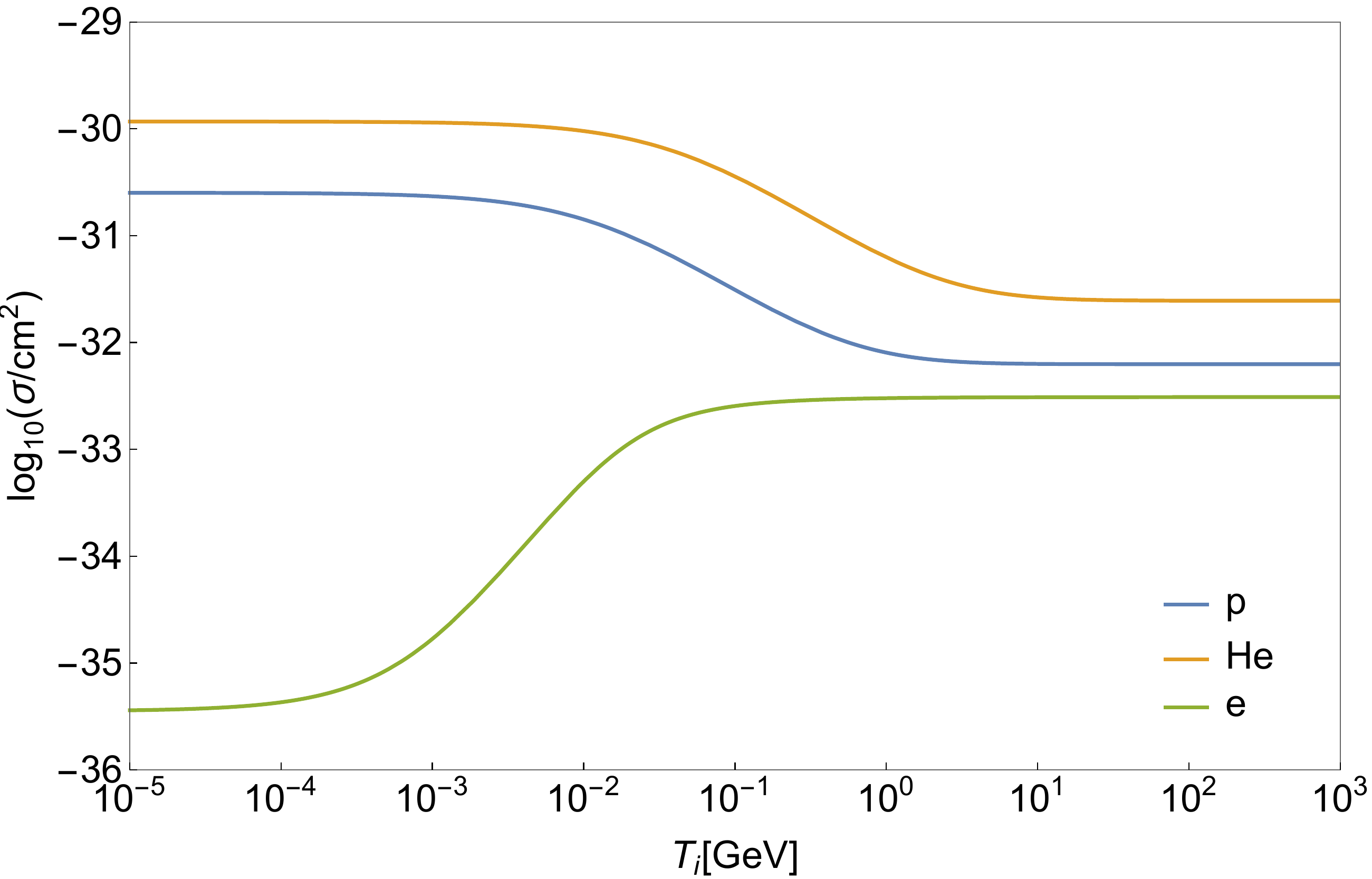}
   \end{tabular}
\end{center}
   \caption{The scattering  cross section of DM and CRs  in the DM rest frame with  the  kinetic energy $T_i$ of CR with DM mass  $m_\chi=10^{-3}\gev$ (Left) and $0.1\gev$ (Right).
  Here we used  $m_{Zd}=0.03\gev$, $\alpha_d=1$, and $\sin^2\varepsilon=10^{-7}$.}
\label{Xection_DMframe}
\end{figure*}

\section{Model}
\label{model}
We consider a model of  Dirac fermion dark matter with a dark gauge symmetry $U(1)_D$ which mediates the interaction between dark and SM sector through mixing with $U(1)_Y$ in the Standard Model~\cite{Essig:2011nj,Baxter:2019pnz,Emken:2019tni}. The  Lagrangian is given by
\begin{equation}
\mathcal{L}_{Z_d}= -\frac{1}{4} \hat{Z}_{d\mu\nu} \hat{Z}_d^{\mu\nu} +\frac{\sin\varepsilon}{2} \hat{B}_{\mu\nu} \hat{Z}_d^{\mu\nu}+\frac12 (m^0_{Z_d})^2\hat{Z}_d^\mu \hat{Z}_{d \mu} ,
\end{equation}
where $\hat{B}_{\mu\nu}$ and $\hat{Z}_{d\mu\nu}$ are the field strengths of $U(1)_{Y}$ in the SM and $U(1)_{D}$ in the dark sector respectively,  with a small mixing term  parametrized by $\sin\varepsilon$, and $m_{Z_d}$ is the mass of dark gauge boson. Here we assume that the hidden sector gauge symmetry is spontaneously broken by additional Higgs so that the mass  of hidden gauge boson $Z_d$ is generated. The fermion dark matter $\chi$ has gauge interaction with hidden gauge boson with gauge coupling $g_d$ as
\begin{equation} 
\mathcal{L}_{int} = g_{d} \hat{Z}_{d\mu} \bar{\chi} \gamma^{\mu} \chi.
\label{int}
\end{equation}

Below the electroweak symmetry breaking, the mass eigenstates (without hat) are related  to the bare gauge fields (with hat) as
\dis{
\hat{A}&= A_{SM} - c_W  t_\varepsilon s_X  Z_{SM} +  c_Wt_\varepsilon  c_X  Z_d,\\
\hat{Z} & = \left(c_X + s_W t_ \varepsilon s_X \right) Z_{SM} + \left( s_X -  s_Wt_\varepsilon c_X \right)Z_d,\\
\hat{Z}_d & = -\frac{s_X}{c_\varepsilon} Z_{SM} + \frac{c_X}{c_\varepsilon}Z_d,
\label{Eq:eigenstates}
}
with the mixing angle $\theta_X$ given by
\begin{equation}
    \tan{2\theta_X}=\frac{2({m_Z^0})^2s_W t_\varepsilon }{({m_Z^0})^2(1-s_W^2 t_\varepsilon^2)-(m_{Z_d}^0)^2/c_\varepsilon^2}.
\end{equation}
Here $m_Z^0$ is the mass of $Z$-boson in the SM, and we use the abbreviations defined by $s_W=\sin\theta_W$, $c_W=\cos\theta_W$ with Weinberg mixing angle $\theta_W$,
and $t_\varepsilon=\tan\varepsilon$, $c_\varepsilon=\cos\varepsilon$,  $s_\varepsilon=\sin\varepsilon$, and similarly for $c_X=\cos\theta_X$, and $s_X=\sin\theta_X$. 

In the SM, the gauge interaction for a fermion $\psi$ with $SU(2)$ charge $T_3$ and electromagnetic charge $Q$ is given by
\dis{
{\mathcal L}_{SM,int} = \bar{\psi}\gamma^\mu\psi \left\{e Q \hat{A}_\mu + \frac{e}{s_W c_W}(T_3-Q s_W^2)\hat{Z}_\mu  \right\},
\label{SMint}
}
where $\psi=\nu_L, e_L, e_R,$ etc and $e=|e|$.
In Appendix, we show the corresponding interaction Lagrangian between DM and proton, neutron, electron and neutrino, from which 
the elastic scattering cross sections are calculated.

For the scattering with nucleus, the cross section at finite momentum transfer is  corrected with a form factor as given by
\dis{
 \sigma_{\chi N} (s,q^2) =  \sigma_{\chi N} (s) \times F^2 (q^2),
}
where $q^2=2m_NT_N$ with the mass of the target $m_N$ and recoil kinetic energy $T_N$.
Here we use the Helm form factor~\cite{Helm:1956zz} with
\dis{
F(q^2)=3\frac{j_1(qr_n)}{q r_n} e^{-q^2 s^2/2},
}
where $j_1$ is the spherical Bessel function, $s=1$ fm is the nuclear skin thickness, and $r_n = (c^2+\frac73\pi^2a^2-5s^2)^{1/2}$ parametrizes the nuclear radius, with $c=1.23 A^{1/3}-0.6$ fm and $a=0.52$ fm, and $A$ is  the mass number of the nucleus.

In Fig.~\ref{Xection_DMframe}, we show the total scattering cross sections in terms of the initial kinetic energy of CRs of proton (blue), He (red), and electron (green), in the rest frame of  DM with mass  $m_\chi=10^{-3}\gev$ (Left) and $0.1\gev$ (Right).  Here we used the parameters  $\mzd=0.03\gev$, $\alpha_d\equiv g_d^2/(4\pi)=1$, and $\sin^2\varepsilon=10^{-7}$. 
We can see that the dependence of the cross section on $T_i$ varies for different mass parameters. When $\mchi < \mzd$ (Left), the cross section is enhances at high $T_i$, however when $\mchi > \mzd$ (Right), the cross section is suppressed at large $T_i$,  due to the relations between the  momentum transfer and the masses of the relevant particles.

\section{Boosted Dark Matter from scatterings with cosmic rays}
\label{BDM}

\begin{figure}[!t]
\begin{center}
\begin{tabular}{cc} 
 \includegraphics[width=0.4\textwidth]{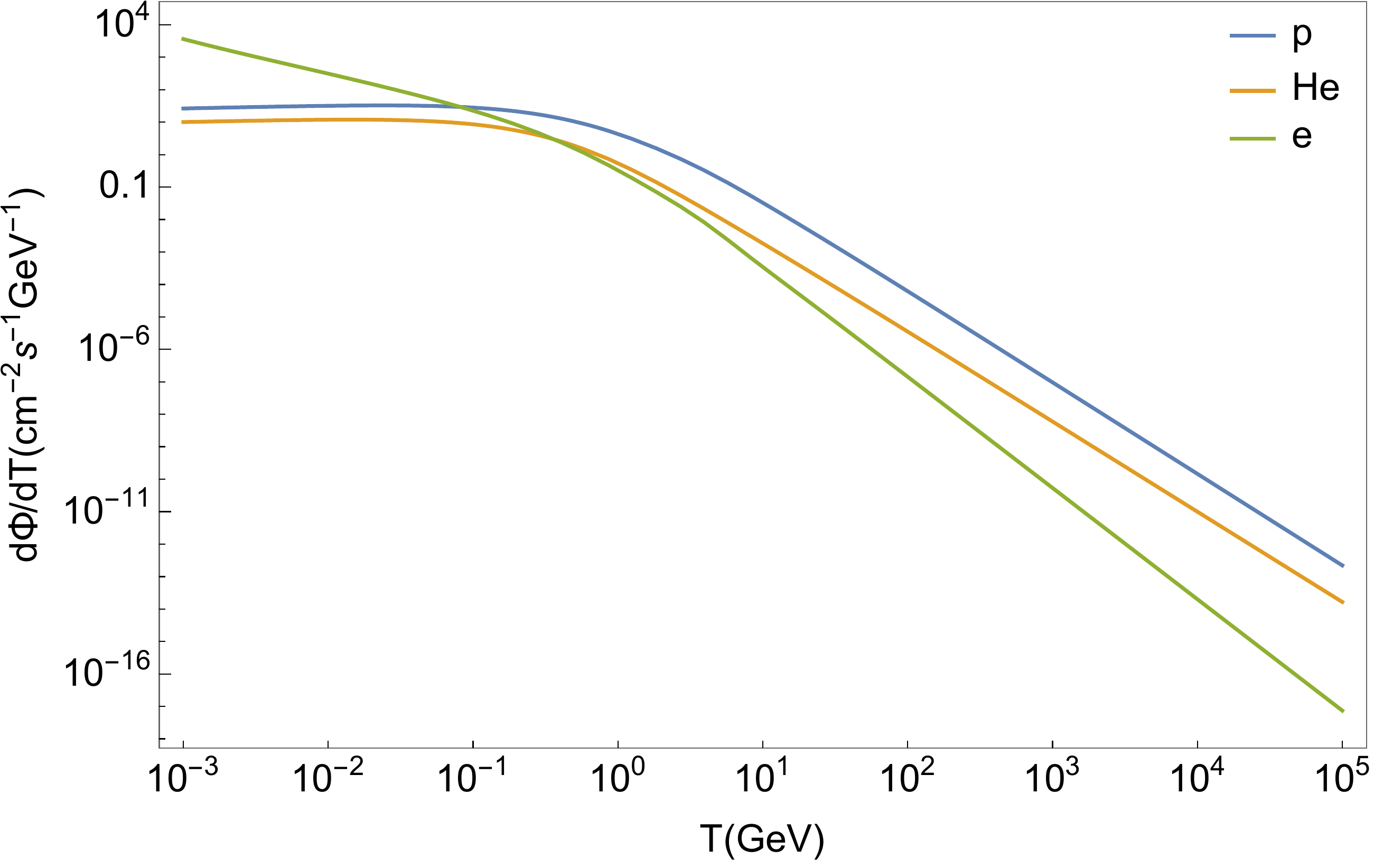}
    \end{tabular}
\end{center}
   \caption{ Differential flux in terms of kinetic energy of CR proton, Helium, and electron~\cite{Bisschoff:2019lne}. }
\label{Flux_CR}
\end{figure}

\begin{figure*}[!t]
\begin{center}
\begin{tabular}{cc} 
 \includegraphics[width=0.4\textwidth]{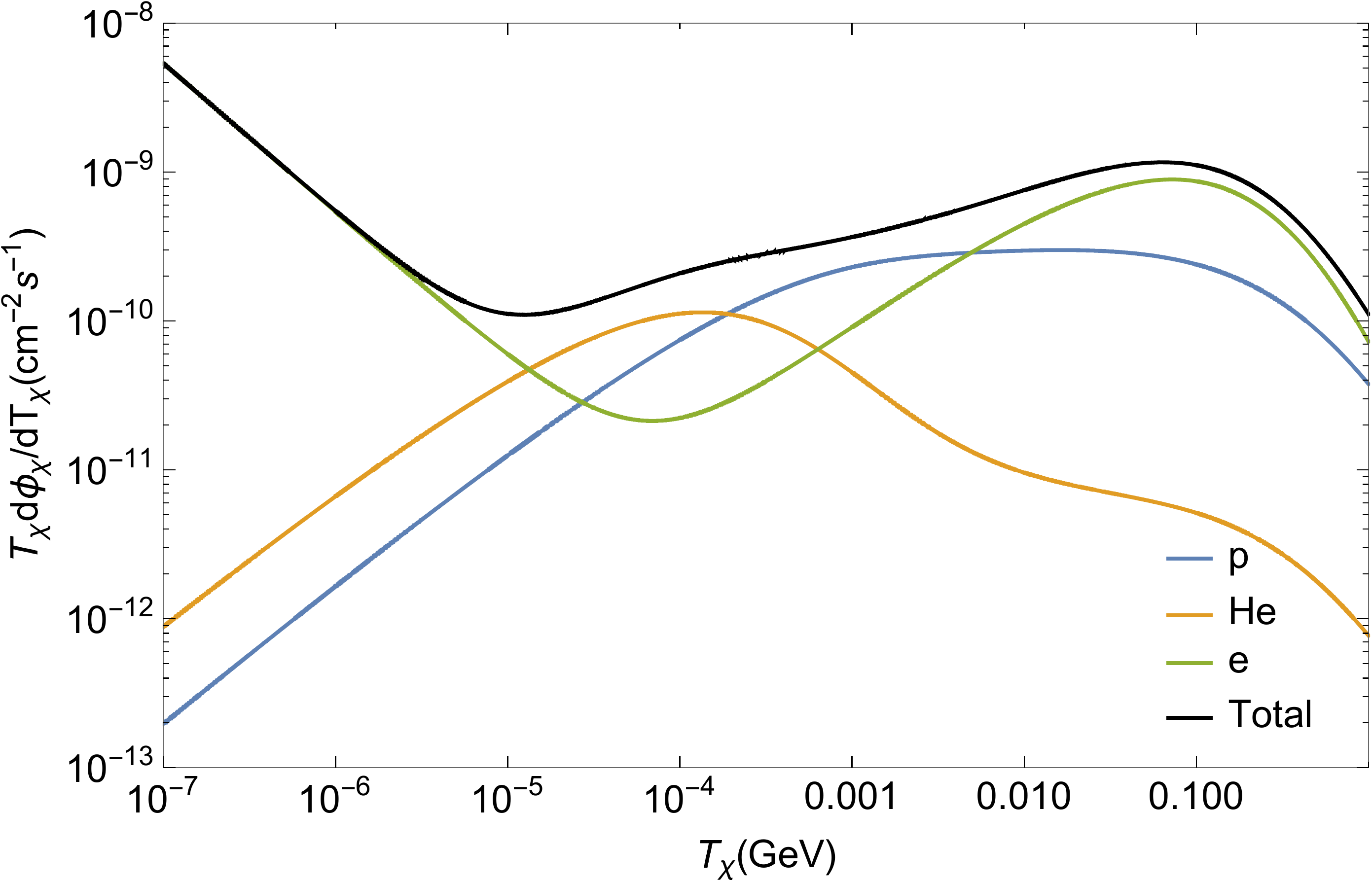}
 &
 \includegraphics[width=0.4\textwidth]{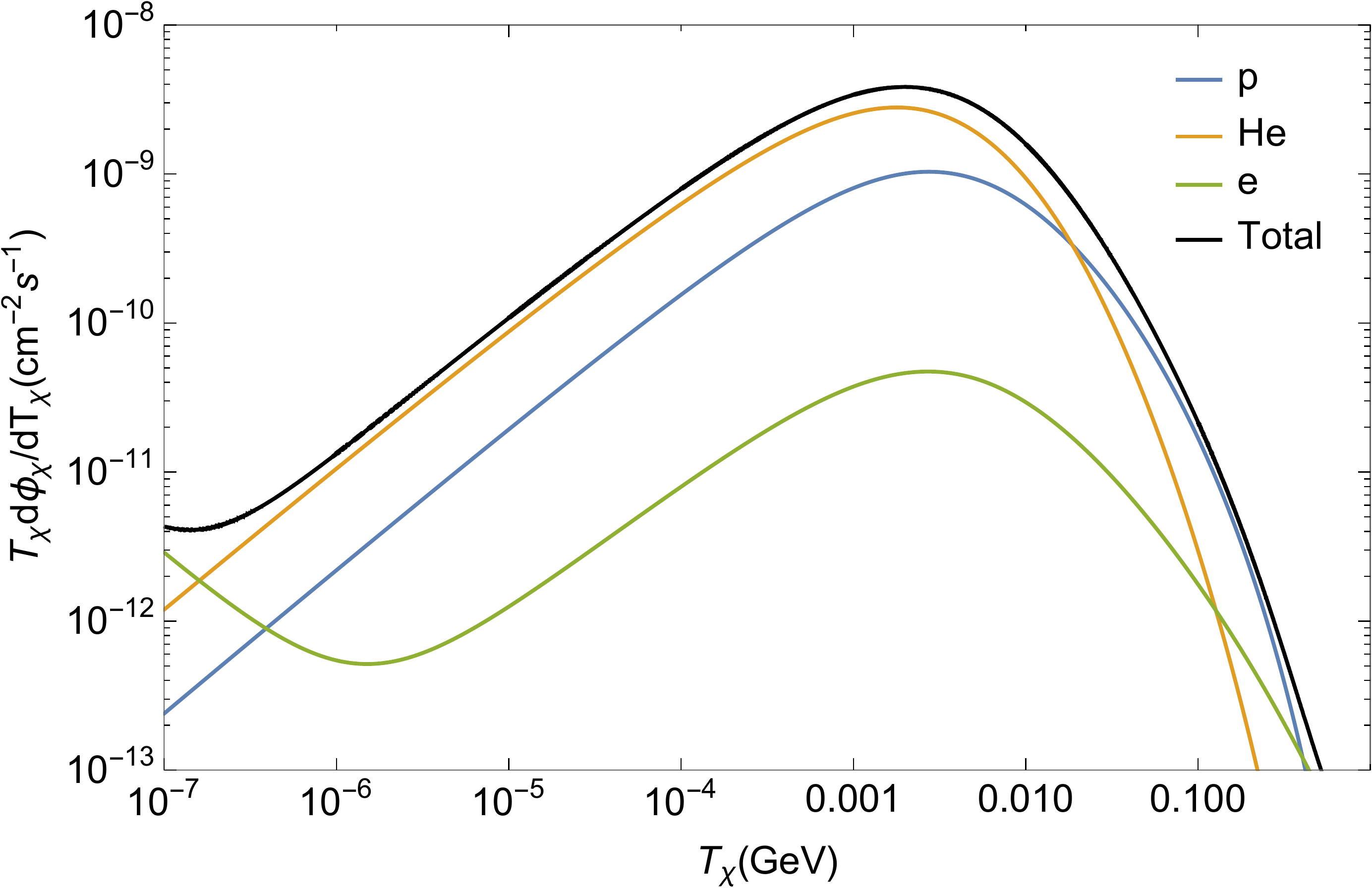}
   \end{tabular}
\end{center}
   \caption{ Flux of BDM around Earth generated from scatterings with proton (blue), He (orange), electron (green), and the total (black).  Here we used $m_\chi=10^{-3}\gev$ (Left), and $0.1\gev$ (Right), with  $m_{Z_d}=0.03\gev$, $\alpha_d=1$, and $\sin^2\varepsilon=10^{-7}$.}
\label{Flux_BDM_darkZ}
\end{figure*}
{\bf Boosted DM} 
The DMs in the Galactic halo are scattered by the cosmic rays. 
In the initial rest frame of DM, the recoiled kinetic energy of DM after scattering $T_\chi$ can be written as  
\begin{equation}
 \label{Tchi}
	\begin{split}
	T_\chi&=T_\chi^{\mathrm{max}}\frac{1-\cos\theta}{2},\\
	T_\chi^{\mathrm{max}}&=\frac{T_i^2+2m_iT_i}{T_i+(m_\chi+m_i)^2/(2m_\chi)},
	\end{split}
\end{equation}
wher $m_{\chi}$ and $m_i$ are the  mass of DM and the colliding CR particle, respectively, and  $\theta$ is  the scattering angle  in the center-of-mass frame between DM and CR particle.
Here $T_\chi^{\mathrm{max}}$ is the maximum kinetic energy that the DM can have after scattering.
The mometum transfer in the collision can be written as $Q^2 = 2m_\chi T_\chi$.
In other way, the minimum kinetic energy of the cosmic particles to make DM with $T_\chi$ is given by
\begin{equation}
	T_i^{\mathrm{min}}=\left(\frac{T_\chi}{2}-m_i\right)\left(1\pm\sqrt{1+\frac{2T_\chi}{m_\chi}\frac{(m_i+m_\chi)^2}{(2m_i-T_\chi)^2}}\right),
\end{equation}
where $+$ for $T_\chi > 2m_i$ and $-$ for  $T_\chi < 2m_i$. When DM collides to the nuclei in the rest frame, $i$ and $\chi$ are interchanged in the  above equations.

To find the flux of BDM, we follow the method in Ref.~\cite{Bringmann:2018cvk}. 
The differential flux of BDM with the kinetic energy $T_\chi$  is obtained by integrating the flux of DM after scattering  with initial kinetic energy of cosmic particle $T_i$ as
\dis{
	\frac{d\Phi_{\chi}}{dT_\chi}=&\sum_{i=p,He,e}\int_{T_i^{\rm min}}^{\infty}\mathrm{d}T_i\, \frac{d\Phi_{\chi}}{dT_\chi dT_i},\\
	=&\,\frac{\rho_\chi^{\rm local}}{m_\chi} D_{\rm eff}  \sum_{i=p,He,e,\nu}\int_{T_i^{\rm min}}^{\infty}\mathrm{d}T_i\, \frac{d\sigma_{\chi i}(T_i)}{dT_\chi} \frac{d\Phi_i^{LIS}}{dT_i},
	\label{dPdTchi}
}
where $T_i^{\rm min}$ is the minimum energy of cosmic rays to give DM kinetic energy $T_\chi$ after collision.
Here we summed over the contributions from each CR of proton, Helium, and electron.
In the second line,  the scattering cross section between DM and CR $\sigma_{\chi i}$ is a function of $T_i$.
For the flux  of cosmic particles, we use the interstellar spectrum of the high energy cosmic particles observed by Voyger 1~\cite{Bisschoff:2019lne}. In Fig.~\ref{Flux_CR}, we show the flux of CRs we used,
 and assume that the CR flux is uniform in the DM halo.

In the second line, the effective distance $D_{eff}$ is defined as
\dis{
D_{\rm eff} = \left(\rho_\chi^{\rm local}\right)^{-1} \int\frac{d\Omega}{4\pi}  \int d\ell \, \rho_\chi,
}
where we used $\rho_\chi^{\rm local}=0.3\gev/\cm^3$. In this paper, as a representative value we use  the effective distance $D_{\rm eff} =  1 \kpc$.

In Fig.~\ref{Flux_BDM_darkZ}, we show the flux  of the BDM generated from scatterings with proton (blue), He (orange), electron (green), and the total (black), for reference values of $m_\chi=0.1\gev$, $m_{Z_d}=30\mev$, $\alpha_d=1$, and $\sin^2\varepsilon=10^{-7}$.
For heavier DM with $\mchi=0.1\gev$ (Right), the proton and Helium dominates, however for the light DM with $\mchi=1\mev$ (Left), the electron scattering is comparable to those from proton and Helium. This can be easily understood from the Fig.~\ref{Xection_DMframe}. When the mass of  DM is lowered, the number of DM increases, and the cross section to nuclei is however decreased at $T_i\sim \gev$, and they more or less compensate. However for electron CR, the cross section is almost the same, and thus the BDM flux increases for lighter DM.  As can be seen from the Fig.~\ref{Flux_BDM_darkZ} (Left) with $\mchi=10^{-3}\gev$,  the contribution of the CR proton and Helium is dominant at $T_\chi \lesssim 0.1\gev $, while the electron contribution is larger at $T_\chi \gtrsim 0.1\gev$.\\

\begin{figure}[!t]
\begin{center}
\begin{tabular}{c} 
 \includegraphics[width=0.4\textwidth]{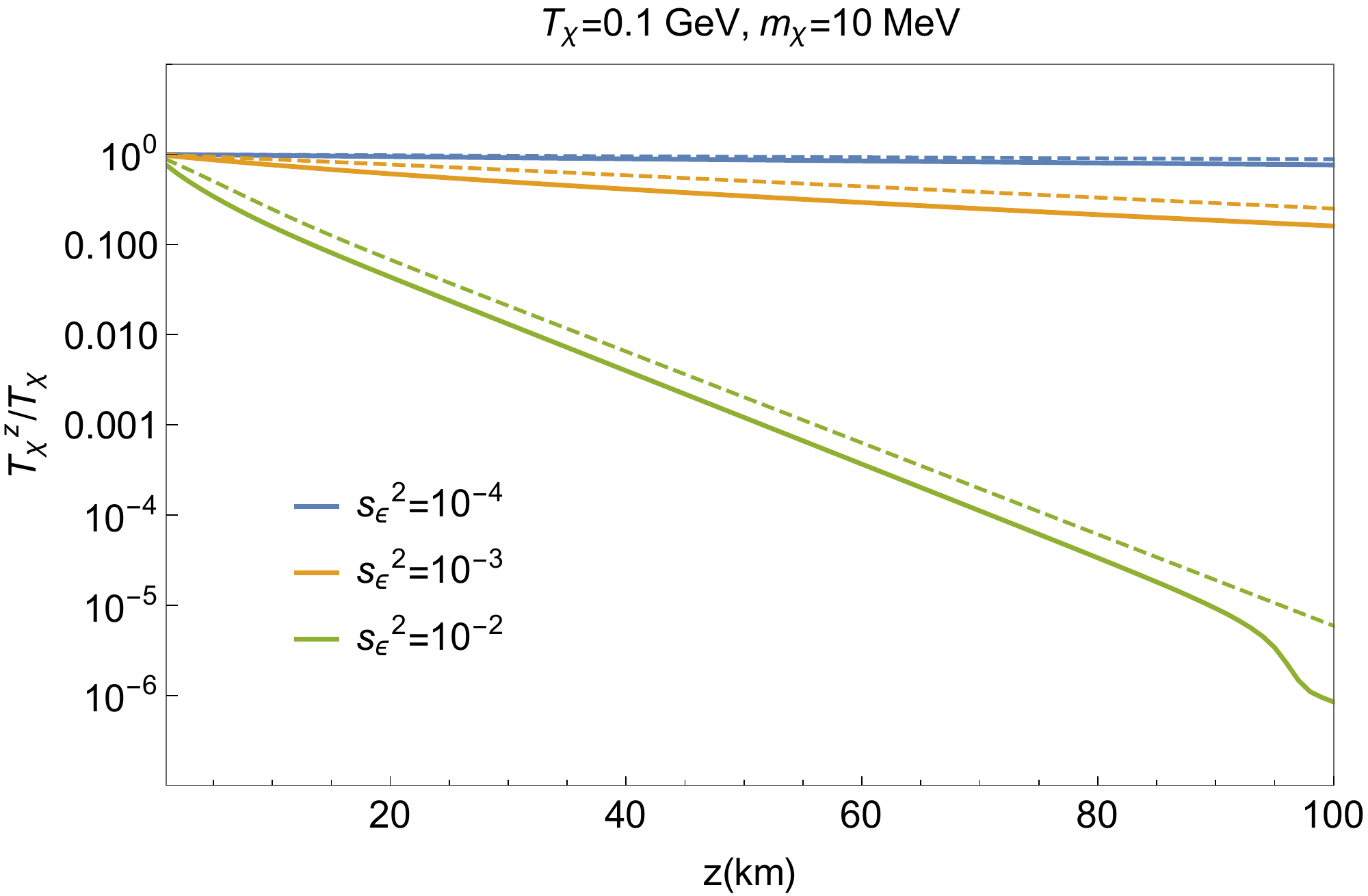}
   \end{tabular}
\end{center}
   \caption{The kinetic energy of BDM  at depth $z$  below the Earth surface  due to the attenuation normalized by the initial kinetic energy. We show them for three cases of the mixing angles $\sin^2\epsilon = 10^{-2}, 10^{-3},10^{-4}$, with  the green, red, and blue lines respectively. The dashed lines show the attenuation with only nuclei ignoring electrons. }
\label{Tchiz}
\end{figure}

{\bf Attenuation}
When the DMs come through the Earth crust, they can interact with the medium and lose energy. This attenuation of  kinetic energy could make DM undetectable because the DMs cannot reach the detector or the kinetic energy of DM become too small for the threshold in the direct detection. The energy loss of DM particles per depth that passing through the medium is~\cite{Bringmann:2018cvk}
\dis{
\label{atten}
	\frac{d T_{\chi}}{d z} & =-\sum_{N} n_{N} \int_{0}^{T_{r}^{\max }} \frac{d \sigma_{\chi N}(T_r)}{d T_{r}} T_{r} \,d T_{r} ,
}
where  $T_r$ is the energy lost by BDM in a collision with nucleus $N$. In a realistic model, the energy dependence of the cross section must be considered. By solving this differential equation, we can find the relation between the DM kinetic energies $T_\chi$ above the Earth and $T_\chi^z$ at depth $z$ below the Earth surface. 
In Fig.~\ref{Tchiz}, we show the change of the kinetic energy of BDM at depth $z$ in the Earth, due to the attenuation with nuclei and electron  (solid) and nuclei alone without electron (dashed) for three cases of mixing with $\sin^2\epsilon = 10^{-2}, 10^{-3},10^{-4}$, with green, red, and blue lines respectively. For low DM mass, the attenuation due to electron is also comparable to that from nuclei, however for heavy DM the attenuation is dominated by the collision with nuclei.

In a recent paper~\cite{Ge:2020yuf}, they studied new diurnal effects from the boosted dark matter due to the anisotropic boosted DM flux and  the Earth attenuation.  However in our paper, the effect is averaged out since we add up the events for a long period to compare with the experiments.

\section{Direct Detection of Boosted DM}
\label{BDMdetection}
The DMs that survived the attenuation of the Earth crust reach the  underground  detector and can scatter the nuclei or the electrons.

\subsection{DM-nucleus interaction}
The  BDMs that reach down the Earth could collide with target nucleus inside in the detector~\cite{Bringmann:2018cvk}. This time, the nucleus is at rest and the DM is moving, which is the opposite situation for upscattering DM by cosmic rays.  The differential  rate per target nucleus is obtained similarly to \eq{dPdTchi} as
\begin{equation} \label{eq:12}
    \frac{d\Gamma_{N}}{dT_{N}} = \int_{T_{\chi}\left(T_{\chi}^{\text{min,z}}\right)}^{\infty} dT_{\chi} \frac{d\sigma_{\chi N}(T_\chi^z)}{d T_{N}} \frac{d \Phi_{\chi}}{d T_{\chi}},
\end{equation}
 where the $T_{\chi}\left(T_{\chi}^{\text {min,z}}\right)$ is  kinetic energy of boosted DM particle outside Earth which gives the minimum kinetic energy to make kinetic energy of target nucleus $T_N$ at the depth $z$  inside the Earth. The scattering cross section is a function of the DM kinetic energy at the location of the detector $\sigma_{\chi N}(T_\chi^z)$, and here $T_\chi^z$ is a function of $T_\chi$ after the attenuation in the Earth, which is evaluated from \eq{atten}.
 
 Then we can calculate the number of the events $N_{\rm sig}$ by integrating between the experimentally accessible recoil energies $T_N\in \{ T_1,T_2  \}$,  for the corresponding observational time $\Delta t$ and target number $N_T$,
 \dis{
 N_{\rm sig} = N_T\times \Delta t \times \int_{T_1}^{T_2}   \frac{d\Gamma_{N}}{dT_{N}},
 }
 and  compare it with the observational constraint.  

For  the present bound, we use the DM search results from a one ton-year exposure of XENON1T~\cite{Aprile:2018dbl}, where  $1.3$ ton Xenon was exposed for 278.8 days with  nuclear recoil energy region between $T_1= 4.9\kev$ and $T_2= 40.9\kev$, and there was no excess found over the background, and thus we  require that  $ N_{\rm sig} < 754$. For future prospect, we use factor 10 higher sensitivity  with Xenon nT~\cite{Aprile:2015uzo}, and 500 to get to the  neutrino floor.

\subsection{DM-electron interaction}
The BDM scatterings with electron can be probed if the recoil energy of the electron $T_e$ is large enough~\cite{Ema:2018bih}. Using the results of  Super-K with 161.9 kton yr~\cite{Kachulis:2017nci}, that is searching signals in the range $0.1\gev < T_e<1.33\gev$ , we apply the number of the events  is smaller than 4042 for 2628.1 days of SK to put the constraint.

\begin{figure}[!t]
\begin{center}
\begin{tabular}{c} 
 \includegraphics[width=0.45\textwidth,angle=0]{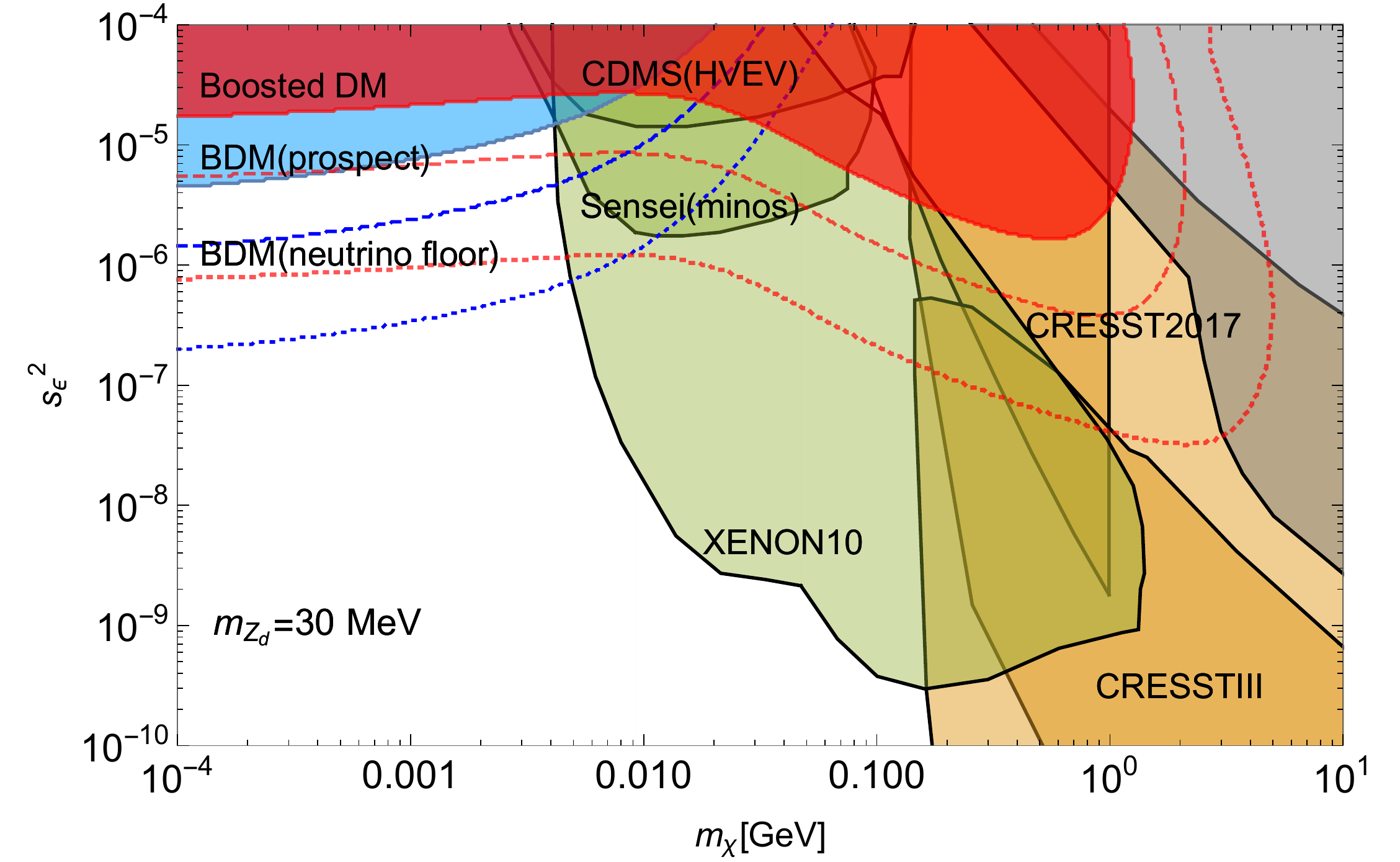}
      \end{tabular}
\end{center}
\caption{Constraints on the DM mass and kinetic mixing from BDM through the scatterings with nuclei (red) and electrons (blue).  Here we used $\alpha_d=1$ and $\mzd=30\mev$. The future prospects are shown with dashed and dotted lines
 The constraints from other direct detection experiments and astrophysical observations  are also shown with thin colors. }
\label{DD_BDM}
\end{figure}

\begin{figure}[!t]
\begin{center}
\begin{tabular}{c} 
 \includegraphics[width=0.45\textwidth]{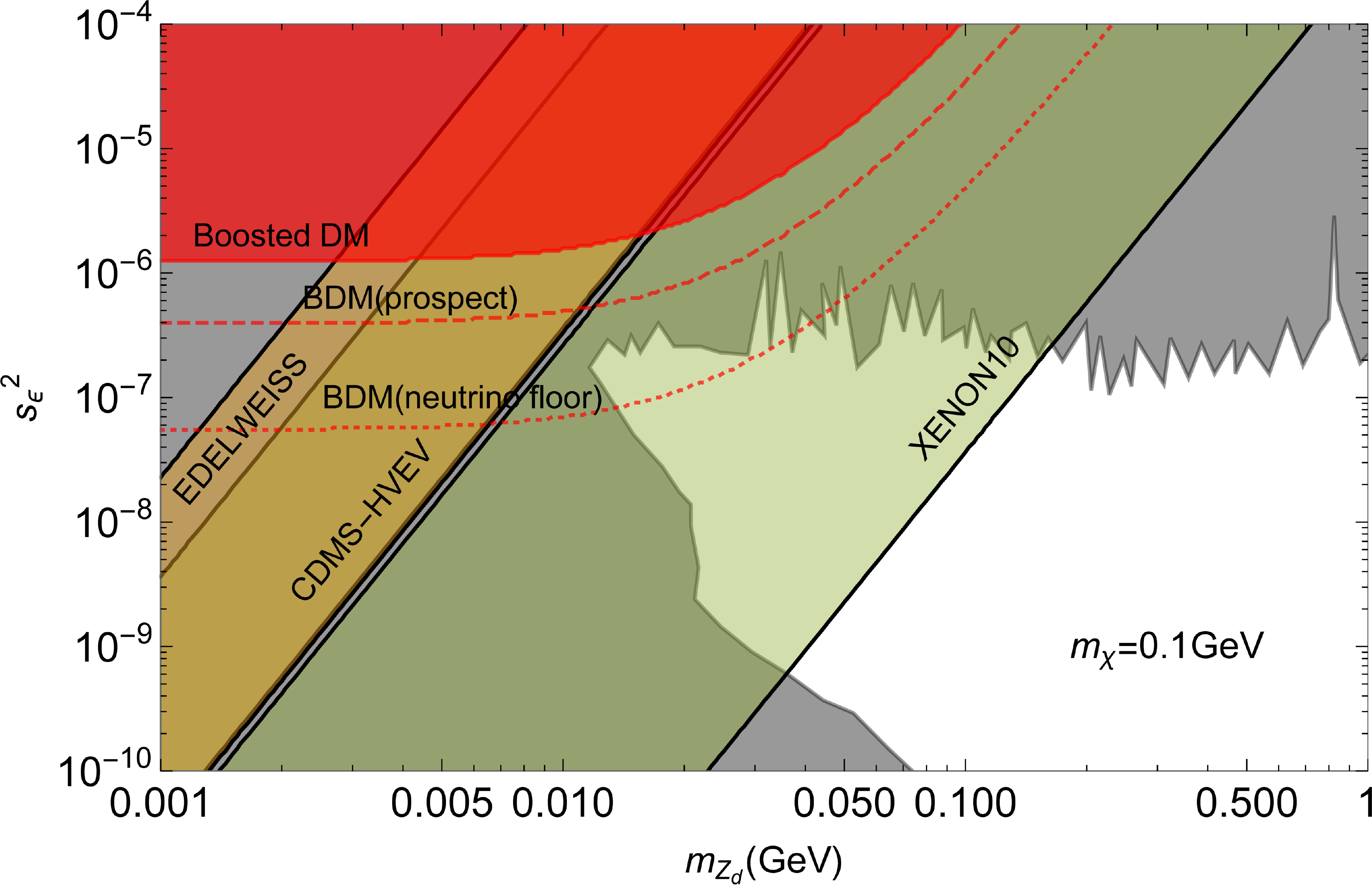}
       \end{tabular}
\end{center}
\caption{Constraints from BDM on the  parameter region ($\mzd,\sin^2\varepsilon^2$) with DM mass $\mchi=100\mev$ for $\alpha_d=1$. Here we used the constraint from Xenon-1T (solid), future prospect (dashed), and neutrino floor (dotted)~\cite{Aprile:2015uzo}. The other constraints shown with thin colors include those from collider, and  beam-dump. The constraints from direct detection are also shown with orange and green colors.
 }
\label{Constraints1}
\end{figure}

\subsection{Results}
In Fig.~\ref{DD_BDM},  we show the constraints on the parameters of $(\mchi,\sin^2\varepsilon)$ from BDM  for  the fixed values of
 $\alpha_d=1$ and $\mzd=30\mev$. The red (blue) shaded region in the left top is disallowed from the direct detection of the BDM with nuclei (electrons) in the detector.  The future prospects are also shown with dashed (10 times) dotted lines (500 times).  The constraints from other experiments are shown with thin colors: direct detection with nuclei (orange)~\cite{Armengaud:2019kfj},  direct detection with electrons (green)~\cite{Emken:2019tni}, and astrophysics and cosmology (grey).

The BDM constraints complements the other bounds of the direct detection with the non-relativistic DM.
This new bound  closes a small open spot at around $\mchi=0.1\gev$ and $\sin^2\varepsilon=3\times 10^{-5}$ and exclude the region of $\mchi < 4\mev$ and $\sin^2\varepsilon\gtrsim 10^{-5}$ which is not probed by non-relativistic DM direct detection. 
However the white region of left-bottom is also constrained when we include the bound from the beam-dump experiments.

In this realistic model of DM, the shape of the constraint is different from those where constant cross section was assumed~\cite{Bringmann:2018cvk,Ema:2018bih,Cappiello:2019qsw}, or that where a simple vector mediation model to the nucleon was used~\cite{Dent:2019krz}.  The $\mchi$-dependence of the BDM constraint can be understood as follows. 

First, the number of DM in the halo is  inversely proportional to $\mchi$.
For the DM-nucleus direct detection, we need to have recoil energy of nucleus larger than keV.  For $\mchi\lesssim 10\mev$, this is satisfied for the DM kinetic energy larger than around 10 MeV, at which the BDM flux is mainly from CR of proton and Helium as well as comparable contribution from electron. 
The energy transferred from CR proton to DM scales as $T_\chi \simeq 2\mchi T_i^2/m_p^2$, and  the integral of the CR flux, which scales $\sim T_i^{-2.7}$,  is proportional to $(T_i^{min})^{-1.7}\propto \mchi^{0.85}$. Therefore the event rate is proportional to $\Gamma \propto \mchi^{-1}\varepsilon^4 \mchi^{0.85} = \varepsilon^4\mchi^{-0.15}$, which gives $\varepsilon^2 \propto \mchi^{0.075}$.  
For $\mchi \gtrsim 10\mev$, the CR proton to DM scattering cross section becomes dependent on $\varepsilon^2\mchi^2$,
and also the recoil energy of nucleus scales $T_N\simeq \frac{2\mchi T_\chi}{m_N}$, with $T\chi\simeq 2\mchi T_i^2/m_p^2$. Therefore 
 $\Gamma \propto \varepsilon^4 \mchi^{2.7}$, which gives $\varepsilon^2 \propto \mchi^{-1.35}$.  That explains the up and down of the BDM constraint (red) in Fig.~\ref{DD_BDM}.

For DM-electron direct detection in Super-K, it is necessary that  the recoil energy of electron be larger than 100 MeV. 
For $\mchi\lesssim 10\mev$, the dominant contribution to BDM comes from CR electron, and for  $\mchi\gtrsim 10\mev$ it comes from CR proton/Helium.
For low $\mchi$ region, the event rate scales as $\Gamma \propto \mchi^{-1}\varepsilon^4$, resulting in $\varepsilon^2\propto\sqrt{\mchi}$. For large $\mchi$ region, $T_e\simeq 2m_e T_\chi^2/\mchi^2$, and $\Gamma \propto \varepsilon^4 \mchi^{-2.7}$. This gives $\varepsilon^2\propto \mchi^{1.35}$~\cite{Ema:2018bih}.

 In Fig.~\ref{Constraints1}, we show the constraints on the plane of $(\mzd,\sin^2\varepsilon)$ for $\mchi=100\mev$ and $\alpha_d=1$,
 with other direct detection bound (orange and green)  as well as the constraints from collider~\cite{Lees:2014xha} and beam-dump experiments~\cite{Alekhin:2015byh} (grey).
 The present BDM constraint is already within the bounds of collider and in the future BDM may touch the unbounded region by them,
 though it is already ruled-out by the Xenon10 experiment.

\section{Astrophysical Constraints}
The large kinetic mixing of the hidden gauge boson with SM may change the effective number of neutrinos, which represents the degrees of freedom of relativistic decoupled species. The current Planck observation gives lower bound on the allowed mass of hidden gauge boson around 8.5 MeV for the mixing parameter $\sin\varepsilon \gtrsim 10^{-9}$~\cite{Ibe:2019gpv}.

The large annihilation of DMs in the early Universe also can affect the BBN and CMB~\cite{Dutra:2018gmv,Krnjaic:2019dzc}. However this may be avoided for a specific models of dark matter such as asymmetric dark matter. This requires non-thermal production of dark matter, which is beyond of our simple model of kinetic mixing~\cite{Baer:2014eja}.

\section{Conclusion}
\label{conclusion}
We studied the impact of the boosted dark matter generated by scatterings of the high energy cosmic rays mediated by the  $U(1)$ gauge kinetic mixing. The non-observation in the underground direct detection combined with the BDM constrains  the light dark matter region, independently of the previous bounds of the direct detection  as well as the collider and beam-dump experiments.

\widetext
\section{Appendix}
\subsection{Kinematics}

The differential cross section for elastic scattering of particle 1 and 2 is given by
\begin{equation}
    \frac{d\sigma}{dt}=\frac{\abs{\overline{\mathcal{M}}}^2}{16\pi \lambda(s,m_1^2,m_2^2)},
    \label{dsdt}
\end{equation}
where $\lambda(a,b,c)=a^2+b^2+c^2-2ab-2bc-2ca=\left[a-\left(\sqrt{b}+\sqrt{c}\right)^2\right]\left[a-\left(\sqrt{b}-\sqrt{c}\right)^2\right]$.\\
If particle 2 is at rest initially, the Mandelstam variables are given by
\begin{equation}
\begin{split}
    s&=m_1^2+m_2^2+2E_1m_2,\\
    &=(m_1+m_2)^2+2T_1m_2=M^2 +2m_1m_2 + 2T_1 m_2,\\
    t&=2m_2^2-2m_2E_2=-2m_2T_2,\\
    u&= 2(m_1^2+m_2^2) - s-t = M^2 -2m_1m_2 -2m_2(T_1-T_2),
        \end{split}
        \label{stu}
\end{equation}
where $M^2= m_1^2+m_2^2$, and 
\dis{
    \lambda(s,m_1^2,m_2^2)&=(s-(m_1+m_2)^2)(s-(m_1-m_2)^2)\\
    &=(2E_1m_2-2m_1m_2)(2E_1m_2+2m_1m_2)\\
    &=4m_2^2(T_1^2+2m_1T_1)\\
    &=2 s \cdot m_2 \cdot T_2^{\mathrm{max}}.
}
Here $T_1$ is the kinetic energy of a particle "1" before collision and $T_2$ is the kinetic energy of a particle "2" after collision, with the maximum of $T_2$ given by\\
\begin{equation}
    T_2^{\mathrm{max}}=\frac{T_1^2+2m_1T_1}{T_1+(m_1+m_2)^2/(2m_2)}.
\end{equation}
Therefore we can write \eq{dsdt} into
\dis{
\frac{d\sigma}{dT_2} =  -2m_2 \frac{d\sigma}{dt} = -\frac{\abs{\overline{\mathcal{M}}}^2}{16\pi s T_2^{\rm max}}.
}
If $\abs{\overline{\mathcal{M}}}^2$ is constant, the total cross section becomes
\begin{equation}
\sigma_{tot}=\int_{-2m_2T_2^{\mathrm{max}}}^0\bfrac{d\sigma}{dt} \mathrm{d}t=\frac{\abs{\overline{\mathcal{M}}}^2}{16\pi s}.
\end{equation}

\subsection{Scattering cross section of DM  in the model of  dark gauge boson}
 The  Lagrangian we are using is written by
\begin{equation}
\mathcal{L} =\mathcal{L}_{\rm SM} -\frac{1}{4} \hat{Z}_{d\mu\nu} \hat{Z}_d^{\mu\nu} +\frac{\sin\varepsilon}{2} \hat{B}_{\mu\nu} \hat{Z}_d^{\mu\nu}+\frac12 ({m^0_{Z_d}})^2\hat{Z}_d^\mu \hat{Z}_{d \mu} + \mathcal{L}_{int},
\end{equation}
where $\hat{B}_{\mu\nu}$ and $\hat{Z}_{d\mu\nu}$ are the field strengths of $U(1)_{Y}$ in the SM and $U(1)_{D}$ in the dark sector respectively,  with a small mixing term  parametrized by $\sin\varepsilon$, and $m_{Z_d}$ is the mass of dark photon.  The fermion dark matter $\chi$ has gauge interaction with hidden gauge boson with gauge coupling $g_d$ as
\begin{equation} 
\mathcal{L}_{int} = g_{d}  \bar{\chi} \gamma^{\mu} \chi \hat{Z}_{d\mu}.
\end{equation}

The mixing term between $\hat{B}$ and $\hat{Z}_d$ can be removed by the field redefinition,
\begin{equation}
	\begin{bmatrix}
	B^0_\mu\\
	\\
	Z^{0}_{d\mu}
	\end{bmatrix}
	=
	\begin{bmatrix}
	1 & -\sin\varepsilon\\
	\\
	0 & \cos\varepsilon
	\end{bmatrix}
	\begin{bmatrix}
	\hat{B}_\mu\\
	\\
	\hat{Z}_{d\mu}
	\end{bmatrix}.
\end{equation}
The electroweak symmetry breaking generates mass to $\hat{Z}$ boson with massless  $\hat{A}$, which are defined by
\dis{
\hat{A}_\mu = c_W \hat{B}_\mu +s_W \hat{W}^3_\mu, \qquad \hat{Z}_\mu =-s_W\hat{B}_\mu+c_W\hat{W}^3_\mu, 
}
in terms of Weinberg mixing angle $\theta_W$ with $c_W\equiv \cos\theta_W$ and $s_W=\sin\theta_W$.
The mass term can be written in terms of $Z^0$ and $Z_d^0$ by
\dis{
\frac12 (m_Z^{0})^2 \hat{Z}_\mu \hat{Z}^{\mu}&=\frac12 (m_Z^{0})^2 (-s_W\hat{B}_\mu+c_W\hat{W}^3_\mu)(-s_W\hat{B}^\mu+c_W\hat{W}^{3,\mu}),\\
& = \frac12 (m_Z^{0})^2 {Z}^0_\mu Z^{0,\mu} - (m_Z^{0})^2 s_W t_\varepsilon  Z^0_\mu Z_d^{0,\mu} + \frac12 (m_Z^{0})^2 s_W^2 t_\varepsilon^2 {Z_d}^0_\mu Z_d^{0,\mu},
}
where
\dis{
Z^0_\mu =-s_WB^0_\mu+c_W\hat{W}^3_\mu,\qquad A^0_\mu =\hat{A}_\mu.
}
Then the mass matrix in the basis of ($A^0$, $Z^0$, $Z_d^0$) is written as
\dis{
M^2 =     \begin{bmatrix}
1& 0&0\\
0& (m_Z^0)^2  & -(m_Z^{0})^2s_W t_\varepsilon \\
0& -(m_Z^{0})^2s_W t_\varepsilon & \frac{(m^0_{Z_d})^2}{\cos^2\varepsilon} +(m_Z^{0})^2 s_W^2 t_\varepsilon^2
\end{bmatrix},
}
which can be diagonalized to find the mass eigenstates ($A_{SM}$, $Z_{SM}$, $Z_d$)
\begin{equation}
    \begin{bmatrix}
    A_{SM\mu}\\
    \\
    Z_{SM\mu}\\
    \\
    Z_{d\mu}
    \end{bmatrix}
    =
    \begin{bmatrix}
    1 & 0 & 0\\
    \\
    0 & \cos{\theta_X} & -\sin{\theta_X}\\
    \\
    0 & \sin{\theta_X} & \cos{\theta_X}
    \end{bmatrix}
    \begin{bmatrix}
    A^0_\mu\\
    \\
    Z^0_\mu\\
    \\
    Z^0_{d\mu}
    \end{bmatrix},
\end{equation}
with the mixing angle $\theta_X$ given by
\begin{equation}
    \tan{2\theta_X}=\frac{2({m_Z^0})^2s_W t_\varepsilon }{({m_Z^0})^2(1-s_W^2 t_\varepsilon^2)-(m_{Z_d}^0)^2/c_\varepsilon^2}.
\end{equation}
The mass eigenvalues for $Z_{SM}$ and $Z_{d}$ are~\cite{Chun:2010ve}
\dis{
m^2_{Z_{SM}} &=(m_Z^0)^2 \left(1+ s_Wt_\varepsilon t_X \right), \\
m^2_{Z_{d}} &= \frac{(m_{Z_d}^0)^2}{c^2_\varepsilon}  \left(1+ s_Wt_\varepsilon t_X \right)^{-1}.\\
}
In this paper, with small $\varepsilon$, we can approximate $m_{Z_{SM}} \simeq m_Z^0$ and $m_{Z_{d}} \simeq m_{Z_d}^0$.
By rearranging the above terms, we can find the relations between the  mass eigenstates of the gauge bosons $(A_{SM}, Z_{SM},Z_d)$ and the interaction eigenstates $(\hat{A}, \hat{Z},\hat{Z}_d)$ as
\dis{
\hat{A}&= A_{SM} - c_W  t_\varepsilon s_X  Z_{SM} +  c_Wt_\varepsilon  c_X  Z_d,\\
\hat{Z} & = \left(c_X + s_W t_ \varepsilon s_X \right) Z_{SM} + \left( s_X -  s_Wt_\varepsilon c_X \right)Z_d,\\
\hat{Z}_d & = -\frac{s_X}{c_\varepsilon} Z_{SM} + \frac{c_X}{c_\varepsilon}Z_d.
\label{Eq:hat}
}
For the standard model, the gauge interaction for a fermion $\psi$ with $SU(2)$ charge $T_3$ and EM charge $Q$ is
\dis{
{\mathcal L}_{int} = \bar{\psi}\gamma^\mu\psi \left\{e Q \hat{A}_\mu + \frac{e}{s_W c_W}(T_3-Q s_W^2)\hat{Z}_\mu  \right\},
}
where $\psi=\nu_L, e_L, e_R,$ etc and $e=|e|$.
By using \eq{Eq:hat}, we can find easily the interaction of SM particles to the mass eigenstates of the gauge bosons.

\subsection{DM-electron scattering}

The interaction Lagrangian of electron is given by
\begin{equation}
\mathcal{L}_{int}=\bar{\bold{e}}\gamma^\mu\bold{e}\left[-eA_{SM\mu}+g_{C}Z_{SM\mu}+g_{Cd}Z_{d\mu}\right]+\bar{\bold{e}}\gamma^\mu\gamma^{5}\bold{e}\left[g_{A}Z_{SM\mu}+g_{Ad}Z_{d\mu}\right],
\end{equation}
where
\begin{equation}
    \begin{split}
        g_{C}&=\frac{e}{4}\left[c_X(3\tan{\theta_W}-\cot{\theta_W})+ \frac{3s_X t_\varepsilon}{ c_W}\right],\\
        g_{Cd}&=\frac{e}{4}\left[s_X(3\tan{\theta_W}-\cot{\theta_W})-\frac{3c _X t_\varepsilon}{ c_W}\right],\\        
        g_{A}&=\frac{e}{4c_W}\left[ \frac{c_X}{ s_W}+ s_X t_\varepsilon  \right],\\
        g_{Ad}&=\frac{e}{4c_W}\left[ \frac{s_X}{s_W} - c_X t_\varepsilon\right].\\
    \end{split}
\end{equation}
Note that $t_X \simeq s_Wt_\varepsilon(1-\mzd^2/m_Z^2)^{-1}$ for very small $\varepsilon$ and $\theta_X$, and thus $g_{Cd}$ and $g_{Ad}$ becomes
\dis{
g_{Cd}\sim& \frac{e}{4} \frac{\mzd^2}{m_Z^2-\mzd^2}\frac{c_W^2-3s_W^2}{c_W}\varepsilon, \\
g_{Ad} \sim &\frac{e}{4}\frac{\mzd^2}{m_Z^2-\mzd^2}\frac{1}{c_W}\varepsilon.
}

The invariant matrix element $\mathcal{M}$ is

\begin{equation}
\begin{split}
i\mathcal{M} & = \bar{u}^s_{p_\chi}\left(ig_{d} \frac{s_{X}}{\sqrt{1-\varepsilon^{2}}} \gamma^{\mu}\right) u^{s'}_{k_\chi}\left[\frac{-i\left(\eta_{\mu \nu}-\frac{q_{\mu} q_{\nu}}{m_{Z}^2}\right)}{q^{2}-m_{Z}^{2}}\right] \bar{u}^r_{p_e}\left(i\gamma^{\nu}(g_{C}+g_{A}\gamma^5)\right) u^{r'}_{k_e} \\
& + 
\bar{u}^s_{p_\chi}\left(-ig_{d} \frac{c_{X}}{\sqrt{1-\varepsilon^{2}}} \gamma^{\mu}\right) u^{s'}_{k_\chi}\left[\frac{-i\left(\eta_{\mu \nu}-\frac{q_{\mu} q_{\nu}}{m_{Z_{d}}^2}\right)}{q^{2}-m_{Z_{d}}^{2}}\right] \bar{u}^r_{p_e}\left(i\gamma^{v} (g_{Cd}+g_{Ad}\gamma^5) \right) u^{r'}_{k_e},
\end{split}
\end{equation}
and the spin-averaged amplitude squared is
\begin{equation}
 \overline{\abs{\mathcal{M}}^2}  = \frac{2 g_{d}^2}{1-\varepsilon^2} \left[\left( \frac{s_{X} g_{C}}{t - m_{Z}^2} - \frac{c_{X} g_{Cd}}{t - m_{Z_{d}}^2} \right)^2 A(m_\chi,m_e) + \left( \frac{s_{X} g_{A}}{t - m_{Z}^2} - \frac{c_{X} g_{Ad}}{t - m_{Z_{d}}^2} \right)^2 B(m_\chi,m_e) \right],
\end{equation}
where

\begin{equation}
\begin{split}
    A(m_\chi,m_i) &= 2t M^2 + (s-M^2)^2 + (u - M^2)^2,\\
    B(m_\chi,m_i) &= (s-M^2)^2 + (u-M^2)^2 + 2t(m_\chi^2 - m_i^2) - 8 m_\chi^2 m_i^2,\\
  {\rm with}\quad    M^2 &= m_\chi^2+m_i^2.\\
\end{split}
\end{equation}
For non-relativistic limit,  $s \rightarrow (m_1+m_2)^2, t \rightarrow 0$, and $ u\rightarrow (m_1-m_2)^2$, then $A(m_\chi,m_i)=8m_\chi^2m_i^2$,  and $B(m_\chi,m_i)=0$.
In this limit,  \eq{stu} becomes
\dis{
t&=-2m_2T_2,\\
s-M^2&=2m_1m_2+2m_2T_1,\\
u-M^2&=-2m_1m_2-2m_2(T_1-T_2).
}
For the non-relativistic limit, $ \overline{\abs{\mathcal{M}}^2}$ becomes
\begin{equation}
   \overline{\abs{\mathcal{M}}^2}  = \frac{16 g_{d}^2 m_\chi^2m_e^2}{1-\varepsilon^2} \left( \frac{s_{X} g_{C}}{m_{Z}^2} - \frac{c_{X} g_{Cd}}{m_{Z_{d}}^2} \right)^2,
\end{equation}
and the scattering cross section is given by
\begin{equation}
    \sigma^{\mathrm{NR}}_{\chi e}=\frac{g_{d}^2 \mu_{\chi e}^2}{\pi(1-\varepsilon^2)}\left(\frac{s_{X} g_{C}}{m_{Z}^2} - \frac{c_{X} g_{Cd}}{m_{Z_{d}}^2}\right)^2.
\end{equation}

\subsection{DM-neutrino scattering}

The interaction Lagrangian of neutrino  is given by
\begin{equation}
    \mathcal{L}_{\text{int}} =  \bar{\bold{\nu}}_e\gamma^\mu(1-\gamma^5)\left[g_{A}Z_{SM\mu}+g_{Ad}Z_{d\mu}\right]\bold{\nu_e}.
\end{equation}.

Withe the invariant matrix element $\mathcal{M}$ given by
\begin{equation}
\begin{split}
i\mathcal{M} & = \bar{\chi}(p^{\prime})\left(ig_{d} \frac{s_{X}}{\sqrt{1-\varepsilon^{2}}} \gamma^{\mu}\right) \chi(p)\left[\frac{-i\left(\eta_{\mu \nu}-\frac{q_{\mu} q_{\nu}}{m_{Z}^2}\right)}{q^{2}-m_{Z}^{2}}\right] \bar{\nu}_{e}\left(k^{\prime}\right)\left(-ig_{A} \gamma^{\nu}(1-\gamma^5)\right) \nu_{e}\left(k\right) \\
& + 
\bar{\chi}(p^{\prime})\left(-ig_{d} \frac{c_{X}}{\sqrt{1-\varepsilon^{2}}} \gamma^{\mu}\right) \chi(p)\left[\frac{-i\left(\eta_{\mu \nu}-\frac{q_{\mu} q_{\nu}}{m_{Z_{d}}^2}\right)}{q^{2}-m_{Z_{d}}^{2}}\right] \bar{\nu}_{e}\left(k^{\prime}\right)\left(-ig_{Ad} \gamma^{v} (1-\gamma^5) \right) \nu_{e}\left(k\right),
\end{split}
\end{equation}
the spin-averaged amplitude squared is obtained as
\begin{equation}
	 \overline{\abs{\mathcal{M}}^2}= \frac{4g_d^2 A(m_\chi,0)}{(1-\varepsilon^{2})} \left(\frac{s_X g_{A}}{(t-m_{Z}^2)}-\frac{c_X g_{Ad}}{(t-m_{Z_{d}}^2)}\right)^2.
\end{equation}

\subsection{DM-nucleus scattering}

The interaction Lagrangian of the proton and neutron is given by
\begin{equation}
\begin{split}
    \mathcal{L}_{int}&=\bar{\bold{p}}\gamma^\mu\bold{p} \left(eA_{SM\mu}-g_{C}Z_{SM\mu}-g_{Cd}Z_{d\mu}\right)+\bar{\bold{p}}\gamma^\mu\gamma^{5}\bold{p}\left(-g_{A}Z_{SM\mu}-g_{Ad}Z_{d\mu}\right)\\
    &+\bar{\bold{n}}\gamma^\mu(1-\gamma^5)\bold{n}\left(-g_{A}Z_{SM\mu}-g_{Ad}Z_{d\mu}\right)
\end{split},
\end{equation}
and thus the interaction of the Nucleus  with mass number $A$ and the number of proton $Z$ is
\begin{equation}
    \mathcal{L}_{int}=\bar{\bold{N}}\gamma^\mu\bold{N}\left[ZeA_{SM\mu}-g_{NC}Z_{SM\mu}-g_{NCd}Z_{d\mu}\right]+\bar{\bold{N}}\gamma^\mu\gamma^{5}\bold{N}\left[-g_{NA}Z_{SM\mu}-g_{NAd}Z_{d\mu}\right],
\end{equation}

\begin{equation}
\begin{split}
    g_{NC}&=Zg_{C}+(A-Z)g_{A},\\
    g_{NCd}&=Zg_{Cd}+(A-Z)g_{Ad},\\
    g_{NA}&=(2Z-A)g_{A},\\
    g_{NAd}&=(2Z-A)g_{Ad}.
\end{split}
\end{equation}

The invariant matrix element $\mathcal{M}$ is

\begin{equation}
\begin{split}
i\mathcal{M} & = \bar{u^s}_{p_\chi}\left(ig_{d} \frac{s_{X}}{\sqrt{1-\varepsilon^{2}}} \gamma^{\mu}\right) u^{s'}_{k_\chi}\left[\frac{-i\left(\eta_{\mu \nu}-\frac{q_{\mu} q_{\nu}}{m_{Z}^2}\right)}{q^{2}-m_{Z}^{2}}\right] \bar{u}^r_{p_N}\left(-i\gamma^{\nu}(g_{NC}+g_{NA}\gamma^5)\right) u^{r'}_{k_N} \\
& + 
\bar{u^s}_{p_\chi}\left(-ig_{d} \frac{c_{X}}{\sqrt{1-\varepsilon^{2}}} \gamma^{\mu}\right) u^{s'}_{k_\chi}\left[\frac{-i\left(\eta_{\mu \nu}-\frac{q_{\mu} q_{\nu}}{m_{Z_{d}}^2}\right)}{q^{2}-m_{Z_{d}}^{2}}\right] \bar{u}^r_{p_N}\left(-i\gamma^{v} (g_{NCd}+g_{NAd}\gamma^5) \right) u^{r'}_{k_N},
\end{split}
\end{equation}
and the spin-averaged amplitude squared is
\begin{equation}
	 \overline{\abs{\mathcal{M}}^2}= \frac{2 g_{d}^2}{1-\var\varepsilon^2} \left[\left( \frac{s_{X} g_{NC}}{t - m_{Z}^2} - \frac{c_{X} g_{NCd}}{t - m_{Z_{d}}^2} \right)^2 A(m_\chi,m_N) + \left( \frac{s_{X} g_{NA}}{t - m_{Z}^2} - \frac{c_{X} g_{NAd}}{t - m_{Z_{d}}^2} \right)^2 B(m_\chi,m_N) \right]
\end{equation}
For non-relativistic limit, it becomes
\begin{equation}
\begin{split}
  \overline{  \abs{\mathcal{M}}^2 } &= \frac{16 g_{d}^2 m_\chi^2m_N^2}{1-\varepsilon^2} \left( \frac{s_{X} g_{NC}}{m_{Z}^2} - \frac{c_{X} g_{NCd}}{m_{Z_{d}}^2} \right)^2,\\
    &= \frac{16 g_{d}^2 m_\chi^2m_N^2}{1-\varepsilon^2} \left( Z\left(\frac{s_{X} g_{C}}{m_{Z}^2} - \frac{c_{X} g_{Cd}}{m_{Z_{d}}^2}\right) + (A-Z)\left(\frac{s_{X} g_{A}}{m_{Z}^2} - \frac{c_{X}g_{Ad}}{m_{Z_{d}}^2}\right) \right)^2,
\end{split}
\end{equation}
and the total scattering cross section becomes
\begin{equation}
    \sigma^{\mathrm{NR}}_{\chi N}=\frac{g_{d}^2 \mu_{\chi N}^2}{\pi(1-\varepsilon^2)} \left( Z\left(\frac{s_{X} g_{C}}{m_{Z}^2} - \frac{c_{X} g_{Cd}}{m_{Z_{d}}^2}\right) + (A-Z)\left(\frac{s_{X} g_{A}}{m_{Z}^2} - \frac{c_{X}g_{Ad}}{m_{Z_{d}}^2}\right) \right)^2.
\end{equation}

{\it Acknowledgments}.
The authors were supported by the National Research Foundation of Korea(NRF) grant funded by the Korea government (MEST) (NRF-2019R1A2B5B01070181). 

\medskip

\bibliographystyle{JHEP}




\end{document}